\documentclass[12pt,a4paper]{article}
\pdfoutput=1

\usepackage[T1]{fontenc} 
\usepackage{amsmath}
\usepackage{amsfonts}
\usepackage{amssymb}
\usepackage{graphicx}
\usepackage{subfig}
\usepackage{hyperref}
\usepackage{color}

\newcommand{\HH}{{\cal H}}

\newcommand{\beq}{\begin{equation}}
\newcommand{\eeq}{\end{equation}}
\newcommand{\bea}{\begin{eqnarray}}
\newcommand{\eea}{\end{eqnarray}}
\def\be{\begin{equation}}
\def\ee{\end{equation}}

\def\beq{\begin{equation}}
\def\eeq{\end{equation}}

\newcommand{\rmd}{\textrm{d}}

\newcommand{\dg}{\delta^{(g)}}
\newcommand{\dgs}{\delta^{(g,s)}}

\begin{document}
\def\thefootnote{\fnsymbol{footnote}}

\begin{center}
\Large{\textbf{Single-Field Consistency Relations of Large Scale Structure \\Part II: Resummation and Redshift Space}} \\[0.5cm]
\end{center}
\vspace{0.5cm}

\begin{center}

\large{Paolo Creminelli$^{\rm a}$, J\'er\^ome Gleyzes$^{\rm b,\rm c}$, Marko Simonovi\'c$^{\rm d, \rm e}$ and
 Filippo Vernizzi$^{\rm b}$}
\\[0.5cm]

\small{
\textit{$^{\rm a}$ Abdus Salam International Centre for Theoretical Physics\\ Strada Costiera 11, 34151, Trieste, Italy}}

\vspace{.2cm}

\small{
\textit{$^{\rm b}$ CEA, Institut de Physique Th\'eorique, F-91191 Gif-sur-Yvette c\'edex, France}}

%\vspace{.2cm}

\small{
\textit{CNRS, Unit\'e de recherche associ\'ee-2306, F-91191 Gif-sur-Yvette c\'edex, France}}
\vspace{.2cm}

\small{
\textit{$^{\rm c}$ Universit\'e Paris Sud, 15 rue George Cl\'emenceau, 91405,  Orsay, France}}

\vspace{.2cm}

\small{
\textit{$^{\rm d}$ SISSA, via Bonomea 265, 34136, Trieste, Italy}}

\vspace{.2cm}

\small{
\textit{$^{\rm e}$ Istituto Nazionale di Fisica Nucleare, Sezione di Trieste, I-34136, Trieste, Italy}}

\vspace{.2cm}

\end{center}

\vspace{.8cm}

\hrule \vspace{0.3cm}
\noindent \small{\textbf{Abstract}\\ 
We generalize the recently derived single-field consistency relations of Large Scale Structure in two directions. First, we treat the effect of the long modes (with momentum $q$) on the short ones (with momentum $k$) non-perturbatively, by writing resummed consistency relations which do not require $k/q \cdot \delta_q \ll 1$. 
These relations do not make any assumptions on the short-scales physics and are extended to include (an arbitrary number of) multiple long modes, internal lines with soft momenta and soft loops.
We do several checks of these relations in perturbation theory and we verify that the effect of soft modes always cancels out in equal-time correlators. 
Second, we write the relations directly in redshift space, without assuming the single-stream approximation: not only the long mode affects the short scales as a homogeneous gravitational field, but it also displaces them by its velocity along the line-of-sight. Redshift space consistency relations still vanish when short modes are taken at equal time: an observation of a signal in the squeezed limit would point towards multifield inflation or a violation of the equivalence principle.}
\\
\noindent
\hrule
\def\thefootnote{\arabic{footnote}}
\setcounter{footnote}{0}

\newpage

\section{Introduction}
Our detailed knowledge of the Universe is mostly based on the study of correlation functions of perturbations around a homogeneous background. A considerable effort has been devoted over the years to the calculation of these correlators during inflation, for the CMB temperature fluctuations and for the present distribution of dark and luminous matter. It is by now well understood that calculations dramatically simplify in the parametric limit in which one (or more) of the momenta (that we call $q$ in this paper) becomes much smaller than the others (denoted by $k$). Recently \cite{Kehagias:2013yd,Peloso:2013zw,Creminelli:2013mca}, these arguments have been applied to the matter (or $\Lambda$)-dominated phase to show that the leading term as $q \to 0$ of any correlation function with $(n+1)$ legs can be written in terms of an $n$-point function: the so-called consistency relations. Although the arguments work in a fully relativistic treatment \cite{Creminelli:2013mca}, which is mandatory if we want to follow the evolution of the modes back in time and connect with inflation, in this paper we focus on the non-relativistic limit, which is valid deep inside the horizon. 

The physical argument behind the consistency relations in the non-relativistic limit is that at leading order in $q$ a long mode gives rise to a homogeneous gravitational field $\vec \nabla\Phi$. The effect of this mode on the short-scale physics can be derived exactly using the equivalence principle and erasing the long mode with a suitable change of coordinates. This logic makes virtually no assumption about the physics at short scales, including the complications due to baryons. However, the cancellation of the long mode by a change of coordinates can be performed only assuming that gravity is all there is: no extra degrees of freedom during inflation (i.e.~single-field inflation) and no extra forces (violation of the equivalence principle) at present. Therefore the consistency relations can be seen as a test of these two assumptions. 

In this paper, which is the natural continuation of \cite{Creminelli:2013mca}, we follow our study of the subject in two directions. First, we want to extend the consistency relations non-linearly in the long mode (Section~\ref{sec:resum}). The displacement due to a homogeneous gravitational field scales with time as $\Delta \vec x \sim \vec \nabla\Phi_L \;t^2$, so that the effect on the short modes of momentum $k$ goes as
\be
\vec  k \cdot  \Delta \vec x \sim k \; q \;\Phi_L  \,t^2 \sim \frac{k}{q} \delta_L \;,
\ee
where $\delta_L$ is the long-mode density contrast\footnote{This is the leading effect in the non-relativistic limit: relativistic corrections are further suppressed by powers of $k/aH \ll 1$ \cite{Creminelli:2013mca}, which are negligible well inside the horizon.}. Notice that this is parametrically larger than $\delta_L$, the natural expansion parameter of perturbation theory, and this is why one is able to capture the leading $q \to 0$ behaviour.  
Obviously, the fact that we can erase a homogeneous gravitational field by going to a free falling frame is an exact statement, that does not require the gravitational field to be small. This implies that we do not need to expand in $k/q \cdot \delta_L$ that can be large, while we keep $\delta_L$ small to allow for a perturbative treatment of the long mode. In Section \ref{sec:resum} we are going to give a resummed version of the consistency relations which is exact in $k/q \cdot \delta_L$. This allows to discuss the case of multiple soft modes and check the relations with the perturbation theory result. With the same logic, we will study the effect of internal soft modes and loops of soft modes. 

The second topic of the paper (Section \ref{sec:redshift}) is to derive consistency relations directly in redshift space, since this is where the distribution of matter is measured. We will do so without assuming anything about the short modes, in particular the single-stream approximation that breaks down in virialized objects. The redshift consistency relations contain an extra piece because the long mode, besides inducing a homogeneous gravitational field in real space, also affects the position of the short modes in redshift space along the line-of-sight. The redshift space consistency relations state that the correlation functions vanish at leading order for $q \to 0$ when the short modes are taken at the same time, as it happens in real space. Given that it is practically impossible, as we will discuss, to study correlation functions of short modes at different times, it is hard to believe that these relations will be verified with real data. However, if a signal is detected at equal times, the consistency relations are not satisfied and this would indicate that at least one of the assumptions does not hold. This would represent a detection of either multi-field inflation or violation of the equivalence principle (or both!).

As explained in \cite{Creminelli:2013mca}, one of the conditions for the validity of the consistency relations is that the long mode has always been out of the sound horizon since inflation. Indeed, a well-understood example where the consistency relations are not obeyed is the case of baryons and cold dark matter particles after decoupling. Before recombination, while dark matter  follows geodesics, baryons are tightly coupled to photons through Thomson scattering and  display acoustic oscillations. Later on, baryons recombine and decouple from photons. Thus, as their sound speed drops they start following geodesics, but with a larger velocity than that of dark matter on comoving scales below the sound horizon at recombination.    As discussed in  \cite{Tseliakhovich:2010bj}, the long-wavelength relative velocity between baryons and CDM  reduces the formation of early structures on small scales, through a genuinely nonlinear effect. 

The fact that baryons have a different initial large-scale velocity compared to dark matter implies, if the long mode is shorter than  the comoving sound horizon at recombination, that the change of coordinates that erases the effect of the long mode is not the same for the two species. Thus the effect of the long mode does not cancel out in the equal-time correlators involving different species  \cite{Bernardeau:2011vy,Bernardeau:2012aq}. In particular, the amplitude of the short-scale equal-time $n$-point functions becomes correlated with the long-wavelength isodensity mode, so that the $(n+1)$-point functions in the squeezed limit do not vanish at equal time. This effect, however, becomes rapidly negligible at low redshifts because the relative comoving velocity between baryons and dark matter decays as the scale factor, $|\vec v_{\rm b} - \vec v_{\rm CDM}| \propto 1/a$.\footnote{The violation of the consistency relations decays as $(D_{\rm iso}/D)^2 \propto (a^2 H f D )^{-2} \sim (1+z)^{3/2}$ where $D_{\rm iso} \propto |\vec v_{\rm b} - \vec v_{\rm CDM}|/( a H f)$ is the growth function of the long-wavelength isodensity mode, $D$ is the growth function of the long-wavelength adiabatic growing mode, $f$ is the growth rate and  $H$ is the  Hubble rate (see \cite{Bernardeau:2011vy} for details); in the last  approximate equality we have used matter dominance. Thus, the effect is already sub-percent at $z \sim 40$.}  Hence, while a deviation can be sizable at high redshifts, it can be neglected in galaxy surveys and the consistency relations apply also when the long mode is shorter than the comoving sound horizon at recombination.
We conclude that the vanishing of the correlation functions at leading order in $q \to 0$ is very robust.

\section{\label{sec:resum}Resumming the long mode}

Let us consider a flat unperturbed FRW universe and add to it a homogenous gradient of the Newtonian potential $\Phi_L$.\footnote{Since we are interested in the non-relativistic limit, we do not consider a constant value of $\Phi_L$, which is immaterial in this limit.} Provided all species feel gravity the same way---namely, assuming the equivalence principle---we can get rid of the effect of $\vec \nabla \Phi_L$ by going into a frame which is free falling in the constant gravitational field. The coordinate change to the free-falling frame is (we are using conformal time $d \eta \equiv dt/a(t)$)
\be
\vec x \to \vec x + \delta \vec x (\eta) \;, \qquad \delta \vec x (\eta) \equiv - \int \vec v_L (\tilde \eta) \, \rmd \tilde \eta\;,  \label{displacement}
\ee
while time is left untouched. The  velocity $\vec v_L$ satisfies the Euler equation in the presence of the homogenous force,
whose solution is 
\be\label{longv}
\vec v_L(\eta) = - \frac{1}{a(\eta)} \int a (\tilde \eta) \vec \nabla \Phi_L (\tilde \eta) \, \rmd\tilde\eta\;.
\ee

To derive the consistency relations we  start from real space. Here, for definiteness, we denote by $\dg$  the density contrast of the galaxy distribution. However, the relations that we will derive are more general and hold for any species---halos, baryons, etc., irrespectively of their bias with respect to the underlying dark matter field. 
Following the argument above, any $n$-point correlation function of short wavelength modes of $\dg$ in the presence of a slowly varying $\Phi_L(\vec y)$ is equivalent to the same correlation function in displaced spatial coordinates, $\vec{\tilde x} \equiv \vec x + \delta \vec x(\vec y, \eta)$, where the displacement field $\delta \vec x(\vec y, \eta)$ is given by eq.~\eqref{displacement} and $\vec y$ is an arbitrary point---e.g., the midpoint between $\vec x_1,  \ldots, \vec x_n$---whose choice is irrelevant at order $q/k$. This statement can be formulated with the following relation,
\be
\begin{split}
\langle \dg (\vec x_1,\eta_1) \cdots \dg (\vec x_n,\eta_n) | {\Phi_L}(\vec y)\rangle &\approx \langle \dg (\vec{\tilde  x}_1,\eta_1) \cdots \dg (\vec{\tilde  x}_n,\eta_n) \rangle_0\; \\
& = \int \frac{\rmd^3k_1}{(2\pi)^3}\cdots\frac{\rmd^3k_n}{(2\pi)^3}  \langle \dg_{\vec k_1} (\eta_1)\cdots \dg_{\vec  k_n} (\eta_n)\rangle_0 \, e^{i \sum_a  \vec k_a \cdot (\vec x_a+  \delta \vec x (\vec y, \eta_a))}  \;, \label{cr1}
\end{split}
\ee
where in the last line we have simply taken the Fourier transform of the right-hand side of the first line. Here and in the following, by the subscript $0$ after an expectation value we mean that the average is taken setting $\Phi_L = 0$ (and not averaging over it); while by $\approx$ we mean an equality that holds in the limit in which there is a separation of scales between long and short modes. 
In momentum space this  holds when the momenta of the soft modes is sent to zero.
In other words, corrections to the right-hand side of $\approx$ are suppressed by  ${\cal O} (q/k)$. 

From eq.~\eqref{displacement} and using the continuity equation $\delta' + \vec \nabla \cdot \vec v =0$, we can rewrite each Fourier mode of the displacement field  as
\be
\delta \vec x(\vec p, \eta) = - i  \frac{\vec p}{p^2} \delta (\vec p, \eta) \equiv - i  \frac{\vec p}{p^2} D(\eta) \delta_0(\vec p) \;, \label{deltax2delta}
\ee
where in the second equality we have defined $D(\eta)$, the growth factor of density fluctuations of the {\em long} mode
and $\delta_0 (\vec p)$, a Gaussian random field with power spectrum $P_0(p)$ which represents the initial condition of the density fluctuations of the long mode \cite{Bernardeau:2001qr}. Notice that the first equality of eq.~\eqref{cr1} is based on the crucial assumption that the long mode is statistically uncorrelated with the short ones. This only works in single-field models of inflation, which we assume throughout. Notice also that eq.~\eqref{deltax2delta}, when going beyond the linear theory, will only receive corrections of order $\delta$, that we can neglect for our purposes since we are only interested in corrections which are enhanced by $1/p$.

At this stage, we can compute an $(n+1)$-point correlation function in the squeezed limit by multiplying the left-hand side of eq.~\eqref{cr1} by $\delta_L$ and averaging over the long mode.
Since the only dependence on $\Phi_L$ in eq.~\eqref{cr1} is in the exponential of $i  \sum_a \vec k_a \cdot \delta \vec x( \vec y,\eta_a)$, we obtain
\be
\label{CoRe1}
\begin{split}
\langle \delta_L( \vec x, \eta) \langle \dg (\vec x_1,\eta_1) \cdots \dg (\vec x_n,\eta_n) | \Phi_L \rangle \rangle_{\Phi_L} \approx &\int \frac{\rmd^3k_1}{(2\pi)^3} \cdots \frac{\rmd^3k_n}{(2\pi)^3}    \langle \dg_{\vec k_1} (\eta_1)\cdots \dg_{\vec  k_n} (\eta_n)\rangle_0 \, e^{i \sum_a \vec k_a \cdot \vec x_a} \\
& \times \int \frac{\rmd^3q}{(2\pi)^3} e^{i  \vec{q}\cdot\vec{x}} \langle\delta_{\vec q} (\eta) e^{i  \sum_a \vec k_a \cdot \delta \vec x( \vec y,\eta_a)}\rangle_{\Phi_L} \, .
\end{split}
\ee
It is then convenient to rewrite this exponential as
\be
\exp \Big[ i  \sum_a \vec k_a \cdot \delta \vec x( \vec y,\eta_a) \Big] =\exp \Big[  \int^\Lambda\frac{\rmd^3p}{(2\pi)^3}  J(\vec p) \delta_0(\vec p) \Big]\;, \label{deltax2J}
\ee
where
\be
J(\vec{p}) \equiv  \sum_a D(\eta_a) \frac{\vec k_a \cdot \vec p}{p^2}   \,e^{i  \vec p \cdot \vec{y}}\, . \label{Jdef}
\ee
The integral is restricted to soft momenta, smaller than a UV cut-off $\Lambda$, which must be much smaller than the hard modes of momenta $k_a$.
Averaging the right-hand side of eq.~\eqref{deltax2J}  over the long wavelength Gaussian random initial condition $\delta_0(\vec p)$ yields\footnote{This result will receive corrections due to primordial non-Gaussianities. Indeed, even in single-field models of inflation, the statistics of modes with comparable wavelength can deviate from Gaussianity. We neglect these corrections in the following.}
\be
\label{SumRes}
\bigg \langle \exp \Big[  \int^{\Lambda}\frac{\rmd^3p}{(2\pi)^3}  J(\vec p) \delta_0(\vec p) \Big] \bigg \rangle_{\Phi_L}  =\exp \bigg[ \frac12 \int^\Lambda\frac{\rmd^3p}{(2\pi)^3}  J(\vec p) J(- \vec p) P_0(p) \bigg]  \;.
\ee
We can use this relation to compute the expectation value of $\delta_L$ with the exponential,
\be
\begin{split}
\bigg \langle \delta_{\vec q}(\eta) \exp \Big( i  \sum_a \vec k_a \cdot \delta \vec x( \vec y,\eta_a) \Big) \bigg \rangle_{\Phi_L}  &= (2\pi)^3 D(\eta) \frac{\delta}{\delta J(\vec q)} \bigg \langle \exp \Big[  \int^\Lambda\frac{\rmd^3p}{(2\pi)^3}  J(\vec p) \delta_0(\vec p) \Big] \bigg \rangle_{\Phi_L}  \\& =  P(q,\eta) \frac{J(- \vec q)}{D(\eta)} 
\exp \bigg[ \frac12 \int^\Lambda\frac{\rmd^3p}{(2\pi)^3}  J(\vec p) J(- \vec p) P_0(p) \bigg] \;,
\end{split}
\ee
where we have defined the power spectrum at time $\eta$: $P( q,\eta) \equiv D^2(\eta) P_0(q)$. 
Finally, rewriting eq.~\eqref{CoRe1} in Fourier space using the above relation and the definition of $J$, eq.~\eqref{Jdef},  we obtain the resummed consistency relations in the squeezed limit,
\be
\begin{split}
\langle  \delta_{\vec q}(\eta) \dg_{\vec k_1} (\eta_1)\cdots \dg_{\vec  k_n} (\eta_n)\rangle' \approx & -   P(q,\eta) \,  \sum_a \frac{D(\eta_a)}{D(\eta)} \frac{\vec k_a \cdot \vec{q}}{q^2}   \;  \langle \dg_{\vec k_1} (\eta_1)\cdots \dg_{\vec  k_n} (\eta_n)\rangle'_0  \\
&  \times \exp \bigg[ {- \frac12 \int^\Lambda \frac{\rmd^3 p}{(2\pi)^3}  \bigg( \sum_a D(\eta_a) \frac{\vec k_a \cdot \vec{p}}{p^2} \, \bigg)^2 P_0(p)} \bigg]  \;, \label{CR}
\end{split}
\ee
where, here and in the following, primes on correlation functions indicate that the momentum conserving delta functions have been removed.
However, what one observes in practice is not the expectation value $\langle\ldots\rangle_0$ with the long modes set artificially to zero: one wants to rewrite the right-hand side of eq.~\eqref{CR} in terms of an average over the long modes. Using eq.~\eqref{SumRes} one gets:
\be
\label{resumloop}
\langle  \langle \dg_{\vec k_1} (\eta_1)\cdots \dg_{\vec  k_n} (\eta_n) | {\Phi_L} \rangle \rangle_{\Phi_L} \approx   \exp \bigg[ {- \frac12 \int^\Lambda \frac{\rmd^3 p}{(2\pi)^3}  \bigg( \sum_a D(\eta_a) \frac{\vec k_a \cdot \vec{p}}{p^2} \, \bigg)^2 P_0(p)} \bigg]   \langle \dg_{\vec k_1} (\eta_1)\cdots \dg_{\vec  k_n} (\eta_n)\rangle_0 \;.
\ee
Once written in terms of the observable quantity the consistency relation comes back to the simple form:
\be
\langle  \delta_{\vec q}(\eta) \dg_{\vec k_1} (\eta_1)\cdots \dg_{\vec  k_n} (\eta_n)\rangle' \approx  -   P(q,\eta) \,  \sum_a \frac{D(\eta_a)}{D(\eta)} \frac{\vec k_a \cdot \vec{q}}{q^2}   \;  \langle \dg_{\vec k_1} (\eta_1)\cdots \dg_{\vec  k_n} (\eta_n)\rangle' \;. \label{CR_easy}
\ee
This equation has the same form as the consistency relations obtained in Refs.~\cite{Kehagias:2013yd,Peloso:2013zw,Creminelli:2013mca}, but now it {\em does not} rely on a linear expansion in the displacement field,
\be
\frac{|\delta \vec x|}{|\vec x|} \sim \frac{k}{q} \delta_L \ll 1 \;.
\ee
Indeed, to derive eq.~\eqref{CR} we have assumed that the long mode is in the linear regime, i.e.~$\delta_L \ll 1$, but no assumption has been made on $({k}/{q}) \delta_L$, which  can be as large as one wishes.
For equal-time correlators the right-hand side vanishes at leading order in $q$ because $\sum_a \vec k_a=\vec q$, in the same way as in the linearized version \cite{Kehagias:2013yd,Peloso:2013zw,Creminelli:2013mca}. 
The resummation of long wavelengths in terms of a global translation of spatial coordinates---whose effect vanishes in equal-time correlation functions---was also performed in  \cite{Bernardeau:2011vy,Bernardeau:2012aq} by using the so-called {\em eikonal} approximation of the equations of motion of standard perturbation theory\footnote{It is not surprising that the consistency relation eq.~\eqref{CR_easy} remains the same even non-linearly in $(k/q) \delta_L$ working directly in terms of the expectation values $\langle\ldots\rangle$ averaged over the long modes. Indeed, neglecting primordial non-Gaussianities, the effect of the mode with momentum $\vec q$ is the same as a change of coordinates, even when the short-scale correlation functions are averaged over all long modes. Since, as we discussed, also eq.~\eqref{deltax2delta} does not require an expansion in $(k/q) \delta_L$, eq.~\eqref{CR_easy} follows.}

It is important to stress that here we made practically no assumptions on the short modes. We did not assume that they are in the linear regime or that the single-stream approximation holds. The relation also takes into account all complications due to baryon physics and it does not assume a description in terms of a Vlasov-Poisson system. We did not assume any model of bias between the short-scale $\dg$ and the underlying dark matter distribution $\delta$. We did not assume that the number of galaxies is conserved at short-scales, so the relation is valid including the formation and merging history. We thus believe that our derivation, rooted only on the equivalence principle, is more robust than the one of  \cite{Kehagias:2013yd,Peloso:2013zw} based on the explicit equations for dark matter and for the galaxy fluid. Notice however that, while we are completely general about the short-modes physics, the long mode is treated in perturbation theory including its bias. Of course what enters in the consistency relations is only the velocity field of the long mode eq.~\eqref{longv}, related to $\Phi_L$ by the Euler equation. In converting this quantity in the density of some kind of objects, one has to rely on the conservation equation and this introduces the issue of bias and of its time-dependence. However, one can measure the large-scale potential in many ways, minimizing the systematic and cosmic-variance uncertainty \cite{Seljak:2008xr}.

As shown below, one can straightforwardly extend this procedure and derive consistency relations involving an arbitrary number of soft legs in the correlation functions or use it to study the effect of soft loops and internal lines.

\subsection{Several soft legs}

The generalisation of the consistency relations above to multiple soft legs (for an analogous discussion in inflation see \cite{Marko}) relies on taking successive functional derivatives with respect to $J(\vec q _i)$ of eq.~\eqref{SumRes}. As an example,  we can explicitly compute the consistency relations with two soft modes.  In this case the $(n+2)$-point function reads
\be
\label{CoRe2}
\begin{split}
 \langle \delta_L( \vec y_1,\tau_1) \delta_L( \vec y_2,\tau_2) \dg (\vec x_1,\eta_1) &\cdots \dg (\vec x_n,\eta_n)  \rangle \approx \int \frac{\rmd^3k_1}{(2\pi)^3} \cdots \frac{\rmd^3k_n}{(2\pi)^3}   \langle \dg_{\vec k_1} (\eta_1)\cdots \dg_{\vec  k_n} (\eta_n)\rangle_0 \,e^{i \sum_a \vec k_a \cdot \vec x_a} \\
& \times \int \frac{\rmd^3q_1}{(2\pi)^3}\frac{\rmd^3q_2}{(2\pi)^3}e^{i  (\vec q_1\cdot\vec y_1+  \vec q_2 \cdot\vec y_2)} \bigg \langle\delta_{\vec q_1}(\tau_1)\delta_{\vec q_2}(\tau_2) e^{\int^\Lambda\frac{\rmd^3p}{(2\pi)^3} J(\vec{p}) \delta_0( \vec p)} \bigg \rangle\, .
\end{split}
\ee
To compute the average over the long modes in the last line, it is enough to take two functional derivatives of eq.~\eqref{SumRes},
\be
\begin{split}
\bigg \langle  \delta_{\vec q_1}(\tau_1)\delta_{\vec q_2}(\tau_2) &e^{\int^\Lambda\frac{\rmd^3p}{(2\pi)^3} J(\vec{p}) \delta_0(\vec p)} \bigg \rangle=(2 \pi)^6D(\tau_1) D(\tau_2) \frac{\delta}{\delta J(\vec q_1)}\frac{\delta}{\delta J(\vec q_2)}\bigg \langle e^{\int^\Lambda\frac{\rmd^3 p }{(2\pi)^3} J(\vec{p}) \delta_0(\vec p)}\bigg \rangle \\
&= \frac{J(-\vec q_1)}{D(\tau_1)}\frac{J(-\vec q_2)}{D(\tau_2)}P(q_1,\tau_1)P(q_2,\tau_2) e^{\frac12 \int^\Lambda \frac{\rmd^3 p}{(2\pi)^3}  J(\vec{p})J(-\vec{p})P_0(p)}\, ,
\end{split}
\ee
where we have assumed $\vec q_1 + \vec q_2 \neq 0$ to get rid of unconnected contributions. In Fourier space, this yields
\be
\label{n+2pf}
\begin{split}
\langle  \delta_{\vec q_1}(\tau_1)\delta_{\vec q_2}(\tau_2) \dg_{\vec k_1} (\eta_1)\cdots \dg_{\vec  k_n} (\eta_n)\rangle' & \approx P(q_1,\tau_1)P(q_2,\tau_2) \\  & \times \sum_a \frac{D(\eta_a)}{D(\tau_1)} \frac{\vec k_a \cdot \vec{q_1}}{q_1^2}    \sum_b \frac{D(\eta_b)}{D(\tau_2)} \frac{\vec k_b \cdot \vec{q_2}}{q_2^2}  \langle \dg_{\vec k_1} (\eta_1)\cdots \dg_{\vec  k_n} (\eta_n)\rangle' \;,
\end{split}
\ee
where again we have used eq.~\eqref{resumloop} to write the result in terms of correlation functions averaged over the long modes.

\begin{figure}
\centering
\includegraphics[width=5in]{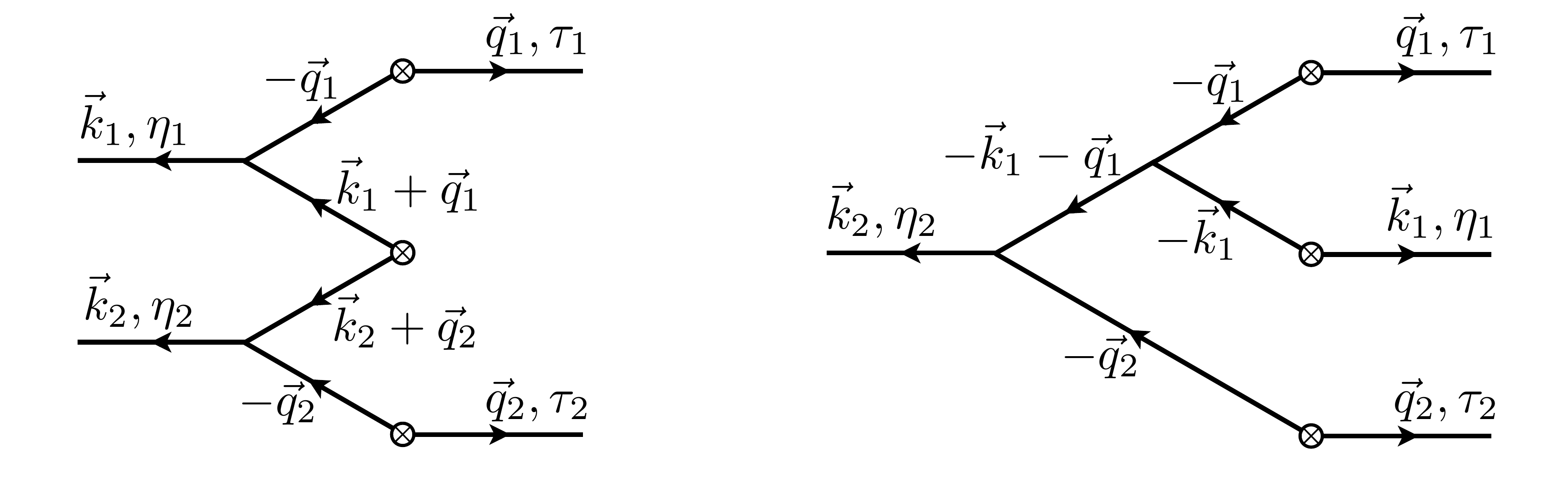}
\caption{\label{fig:trispectrum} \small{Two diagrams that contribute to the tree-level trispectrum. Left: $T_{1122}$. Right: $T_{1113}$.}}
\end{figure}

As a simple example, let us consider eq.~\eqref{n+2pf} in the case where $n=2$ and $\dg$ describes dark matter perturbations, i.e.~$\dg \equiv \delta$.
In this case, at lowest order in $\frac k q \delta(\vec q,\eta)$---i.e.~setting the exponential in the third line to unity---the above relation reduces to
\be
\label{TriSpec}
\langle\delta_{\vec q_1}(\tau_1) \delta_{\vec q_2}(\tau_2) \delta_{\vec k_1}(\eta_1) \delta_{\vec k_2}(\eta_2)\rangle' \approx \frac{(D(\eta_1) - D(\eta_2) )^2}{D(\tau_1) D(\tau_2) } \frac{\vec q_1\cdot \vec k_1}{q_1^2} \frac{\vec q_2\cdot \vec k_1}{q_2^2}P(q_1,\tau_1)P(q_2,\tau_2) \langle \delta_{\vec k_1}(\eta_1)\delta_{\vec k_2}(\eta_2)\rangle'.
\ee
We can check that this expression correctly reproduces the tree-level trispectrum computed in perturbation theory in the double-squeezed limit.  This can be easily computed by summing the two types of diagrams displayed in Fig.~\ref{fig:trispectrum}.
The diagram on the left-hand side represents the case where the density perturbations of the short modes are both taken at second order, yielding
\be
\label{cont_t_1}
\begin{split}
T_{1122}=& \, D(\tau_1) D(\tau_2) D(\eta_1) D(\eta_2)  P_{0}(q_1)P_{0}(q_2)  F_2(-\vec q_1, \vec k_1+\vec q_1) F_2(-\vec q_2, \vec k_2+\vec q_2)  \langle \delta_{\vec k_1} (\eta_1)\delta_{\vec k_2} (\eta_2)\rangle' + {\rm perms} \\
& \approx -8 \frac{\vec q_1\cdot \vec k_1}{2 q_1^2} \frac{\vec q_2\cdot \vec k_1}{2 q_2^2} \frac{D(\eta_1) D(\eta_2)}{D(\tau_1) D(\tau_2)} P(q_1,\tau_1)P(q_2,\tau_2) \langle \delta_{\vec k_1}(\eta_1)\delta_{\vec k_2}(\eta_2)\rangle' \, ,
\end{split}
\ee
where, on the right-hand side of the first line, $ F_2(\vec p_1,\vec p_2)$ is the usual kernel of perturbation theory, which in the limit where $p_1 \ll p_2$ simply reduces to $ \vec p_1 \cdot \vec p_2/( 2 p_1^2)$  \cite{Bernardeau:2001qr}.
The second type of diagram, displayed on the right-hand side of Fig.~\ref{fig:trispectrum}, is obtained when one of the short density perturbations is taken at third order; it gives
\be
\label{cont_t_2}
\begin{split}
T_{1113} &= D(\eta_2)^2 D(\tau_1) D(\tau_2) P_{0}(q_1)P_{0}(q_2) F_3(-\vec q_1, - \vec q_2, - \vec k_1) \langle \delta_{\vec k_1}(\eta_1)\delta_{\vec k_2}(\eta_2)\rangle' + {\rm perms} \\
& \approx   4 \frac{\vec q_1\cdot \vec k_1}{2 q_1^2} \frac{\vec q_2\cdot \vec k_1}{2 q_2^2} \frac{D(\eta_2)^2}{D(\tau_1) D(\tau_2)} P(q_1,\tau_1)P(q_2,\tau_2) \langle \delta_{\vec k_1}(\eta_1)\delta_{\vec k_2}(\eta_2)\rangle' \, ,
\end{split}
\ee
where, on the right-hand side of the first line, $ F_3(\vec p_1,\vec p_2,\vec p_3)$ is the third-order  perturbation theory kernel,  which in the limit where $p_1, p_2 \ll p_3$  reduces to $ (\vec p_1 \cdot \vec p_3)(\vec p_2 \cdot \vec p_3) /( 4 p_1^2 p_2^2 )$ \cite{Bernardeau:2001qr}.
As expected, summing up all the contributions to the connected part of the trispectrum, i.e.~$T_{1122}+T_{1131}+T_{1113}$, using eqs.~\eqref{cont_t_1} and \eqref{cont_t_2} and $\vec k_2 \approx - \vec k_1$ one obtains eq.~\eqref{TriSpec}.

\subsection{Soft Loops}

So far we have derived  consistency relations where the long modes appear explicitly as external legs. We now show that our arguments can also capture the effect on short-scale correlation functions of soft modes running in loop diagrams.
We already did this in eq~\eqref{resumloop}
\be
\langle  \langle \dg_{\vec k_1} (\eta_1)\cdots \dg_{\vec  k_n} (\eta_n) | {\Phi_L} \rangle \rangle_{\Phi_L} \approx   \exp \bigg[ {- \frac12 \int^\Lambda \frac{\rmd^3 p}{(2\pi)^3}  \bigg( \sum_a D(\eta_a) \frac{\vec k_a \cdot \vec{p}}{p^2} \, \bigg)^2 P_0(p)} \bigg]   \langle \dg_{\vec k_1} (\eta_1)\cdots \dg_{\vec  k_n} (\eta_n)\rangle_0 \;.
\ee
The exponential in this expression can be expanded at a given order, corresponding to the number of soft loops dressing the $n$-point correlation function. Each loop  carries a contribution $\propto k^2 \int \rmd p P_0(p) $ to the correlation function. However, 
this expression makes it very explicit that at all loop order these contributions have no effect on equal-time correlators, because in this case the exponential on the right-hand side is identically unity. This  confirms previous analysis on this subject  \cite{Jain:1995kx,Scoccimarro:1995if,Bernardeau:2011vy,Bernardeau:2012aq,Blas:2013bpa,Carrasco:2013sva}. It is important to notice again, however, that in our derivation this cancellation is more general and robust that in those references, as it takes place independently of the equations of motion  for the short modes and  is completely agnostic about the short-scale physics. It simply derives from the equivalence principle.

Nevertheless, soft loops contribute to unequal-time correlators. As a check of the expression above, one can compute the contribution of soft modes to the 1-loop unequal-time matter power spectrum, $\langle \delta_{\vec k_1}(\eta_1) \delta_{\vec k_2}(\eta_2)\rangle'$, and verify that this reproduces the standard perturbation theory result. Expanding at order $( \frac kp \delta)^2$ the exponential in eq.~\eqref{resumloop} for $n=2$, one obtains the 1-loop contribution to the power spectrum,
\be
\label{1loop}
\langle \dg_{\vec k}(\eta_1) \dg_{-\vec k}(\eta_2)\rangle'_{\rm 1-soft\,loop}\approx -\frac12\left({D(\eta_1)-D(\eta_2)}\right)^2 \int^\Lambda \frac{\rmd^3 p}{(2\pi)^3}\bigg(\frac{\vec p \cdot \vec k}{ p^2}\bigg)^2  P_0(p) \langle \dg_{\vec k}(\eta_1) \dg_{-\vec k}(\eta_2)\rangle'_0 \, .
\ee
Let us now compute the analogous contribution in perturbation theory. 
\begin{figure}
\centering
\includegraphics[width=5in]{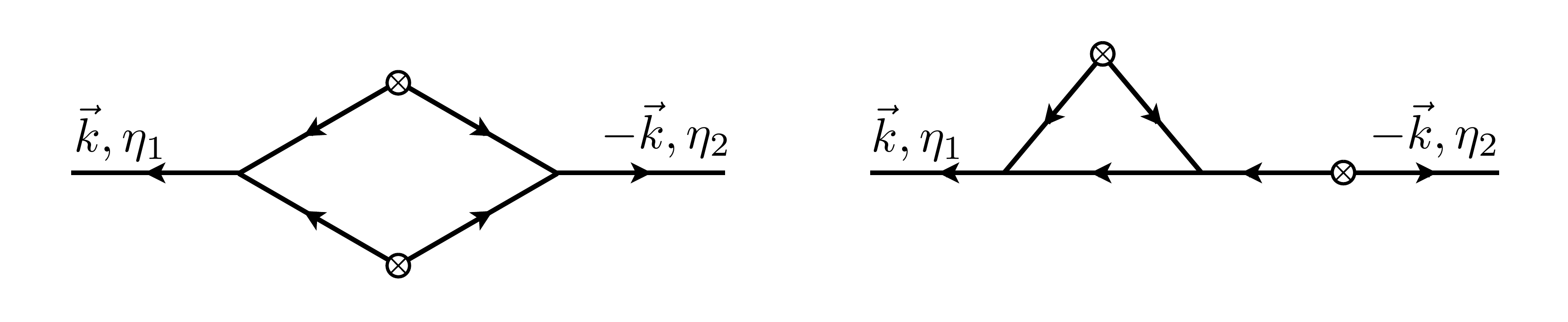}
\caption{\label{fig:loop} \small Two diagrams that contribute to the 1-loop power spectrum. Left: $P_{22}$. Right: $P_{31}$.}
\end{figure}
Two types of diagrams are going to be relevant; these are shown in Fig.~\ref{fig:loop}. The one on the left, usually called $P_{22}$, yields
\be
P_{22}\approx  4 D(\eta_1) D(\eta_2) \int^\Lambda \frac{\rmd^3 p}{(2\pi)^3}\bigg(\frac{\vec p \cdot \vec k}{2 p^2}\bigg)^2  P_0(p) \langle \delta_{\vec k}(\eta_1) \delta_{-\vec k}(\eta_2)\rangle'_0 \, ,
\ee
while the diagram on the right, $P_{31}$,  gives
\be
P_{31} \approx - 2 D(\eta_1)^2 \int^\Lambda \frac{\rmd^3 p}{(2\pi)^3}\bigg(\frac{\vec p \cdot \vec k}{2 p^2}\bigg)^2 P_0(p)  \langle \delta_{\vec k}(\eta_1) \delta_{-\vec k}(\eta_2)\rangle'_0 \, .
\ee
Summing up all the different contributions, $P_{22} + P_{31} + P_{13}$, one obtains eq.~\eqref{1loop}.

\subsection{Soft internal lines}
Another kinematical regime in which the consistency relations can be applied is the limit in which the sum of some of the external momenta becomes very small, for instance $|\vec k_1 +\cdots + \vec k_m |\ll k_1, \ldots, k_m$. In this limit, the dominant contribution to the $n$-point function comes from the diagram where $m$ external legs of momenta $\vec k_1,\ldots,\vec k_m$ exchange soft modes with momentum $\vec q=\vec k_1 +\cdots + \vec k_m$ with $n-m$ external legs with momenta $\vec k_{m+1}, \ldots, \vec k_n$ (for an analogous case in inflation see \cite{Seery:2008ax,Leblond:2010yq}). In the language of our approach, this contribution comes from averaging a product of $m$-point and $(n-m)$-point functions under the effect of long modes. 

\def\dg{\delta}

In this case, the $n$-point function in real space can be written as
\be
\begin{split}
\langle\dg(\vec x_1,\eta_1) & \cdots \dg(\vec x_m, \eta_m)  \; \dg(\vec x_{m+1},\eta_{m+1}) \cdots \dg(\vec x_n,\eta_n) \rangle  \\
& \approx \langle \langle\dg(\vec x_1,\eta_1) \cdots \dg(\vec x_m, \eta_m) | \Phi_L \rangle \langle \dg(\vec x_{m+1},\eta_{m+1}) \cdots \dg(\vec x_n,\eta_n) | \Phi_L \rangle \rangle_{\Phi_L}\;,
\end{split}
\ee
where here and in the rest of the section we drop the superscript ${}^{(g)}$ on the galaxy density contrast to lighten the notation.
Now we can straightforwardly apply the equations from the previous sections. As before, the long mode can be traded for the change of coordinates. Rewriting the right-hand side in Fourier space we get 
\be
\label{isl1}
\begin{split}
\langle\dg(\vec x_1,\eta_1) & \cdots \dg(\vec x_m,\eta_m)\; \dg(\vec x_{m+1},\eta_{m+1}) \cdots \dg(\vec x_n,\eta_n)\rangle  \\
& \approx \int \frac{\rmd^3k_1}{(2\pi)^3}\cdots\frac{\rmd^3k_n}{(2\pi)^3} \langle\dg_{\vec k_1}(\eta_1)\cdots \dg_{\vec k_m}(\eta_m) \rangle_0 \langle \dg_{\vec k_{m+1}}(\eta_{m+1}) \cdots \dg_{\vec k_n}(\eta_n)\rangle_0  \, e^{i \sum_a  \vec k_a \cdot \vec x_a} \\
& \times \left \langle \exp\bigg[i  \sum_{a=1}^m \vec k_a \cdot \delta \vec x( \vec y_1,\eta_a)\bigg] \cdot \exp\bigg[i \sum_{a=m+1}^n \vec k_a \cdot \delta \vec x( \vec y_2,\eta_a)\bigg]  \right \rangle_{\Phi_L} \;,
\end{split}
\ee
where $\vec y_1$ and $\vec y_2$ are two different points  respectively close to $(\vec x_1,\vec x_2, \ldots, \vec x_m )$ and $(\vec x_{m+1}, \vec x_{m+2} , \ldots ,\vec x_n)$. 
The average over the long mode can be rewritten as
\be
\left \langle \exp\left[\int^\Lambda \frac{\rmd ^3 \vec p}{(2\pi)^3} \big( J_1(\vec p)+ J_2(\vec p) \big) \delta_0(\vec p) \right] \  \right \rangle_{\Phi_L}
\ee
with
\be
J_1(\vec p) = \sum_{a=1}^m D(\eta_a) \frac{\vec k_a \cdot \vec p}{p^2} e^{i \vec p \cdot \vec y_1} \;, \quad J_2(\vec p) = \sum_{a=m+1}^n D(\eta_a) \frac{\vec k_a \cdot \vec p}{p^2} e^{i \vec p \cdot \vec y_2} \;.
\ee
Taking the expectation value over the long mode using the  expression for averaging the exponential of a Gaussian variable, i.e.~eq.~\eqref{SumRes},  eq.~\eqref{isl1} can be written as
\be
\label{sl2}
\begin{split}
\langle\dg(\vec x_1,\eta_1) & \cdots \dg(\vec x_m,\eta_m)\; \dg(\vec x_{m+1},\eta_{m+1}) \cdots \dg(\vec x_n,\eta_n)\rangle  \\
& \approx  \int \frac{\rmd^3k_1}{(2\pi)^3}\cdots\frac{\rmd^3k_n}{(2\pi)^3}\langle\dg_{\vec k_1}(\eta_1)\cdots \dg_{\vec k_m}(\eta_m) \rangle ' \langle \dg_{\vec k_{m+1}}(\eta_{m+1}) \cdots \dg_{\vec k_n}(\eta_n)\rangle'   \, e^{i \sum_a \vec k_a \cdot \vec x_a}\\
& \times \exp \left[ - \int^\Lambda \frac{\rmd^3 p}{(2\pi)^3} J_1(\vec p) J_2(\vec p) P_0(\vec p) \right]  \;.
\end{split}
\ee

We are interested in the soft internal lines, that come from the cross term, i.e.~the last line of eq.~\eqref{sl2}. Notice that $J_1(\vec p)$ and $J_2(\vec p)$ are evaluated at different points $\vec y_1$ and $\vec y_2$ separated by a distance $\vec x$.\footnote{For definiteness, we can choose $\vec y_1=\frac{1}{m}\sum_{a=1}^m \vec x_a$ and $\vec y_2=\frac{1}{n-m}\sum_{a=m+1}^n \vec x_a$.}
It is lengthy but straightforward to take the Fourier transform of this equation, which yields
\be
\begin{split}
\label{soft_lines_final}
\langle\dg_{\vec k_1}(\eta_1)& \cdots \dg_{\vec k_m}(\eta_m) \dg_{\vec k_{m+1}}(\eta_{m+1})\cdots \dg_{\vec k_n}(\eta_n) \rangle' \\ &
\approx \langle\dg_{\vec k_1}(\eta_1)\cdots \dg_{\vec k_m}(\eta_m) \rangle' \,  \langle \dg_{\vec k_{m+1}}(\eta_{m+1}) \cdots \dg_{\vec k_n}(\eta_n)\rangle'\\
&\times\int \rmd^3  x  \, e^{-i\sum_{i=1}^m \vec k_i\cdot \vec x} \exp \bigg[ - \int^\Lambda \frac{\rmd ^3  p}{(2\pi)^3}e^{i\vec p\cdot \vec x}   \, \sum_{a=1}^{m} D(\eta_a) \frac{\vec k_a \cdot \vec{p}}{p^2}  \sum_{a=m+1}^{n} D(\eta_a) \frac{\vec k_a \cdot \vec{p}}{p^2} \, P_0(p)\bigg]\,.
\end{split}
\ee
The last line  encodes the effect of soft modes with total momentum $\vec q=\vec k_1 +\cdots + \vec k_m$ exchanged between $m$ external legs of momenta $\vec k_1,\ldots,\vec k_m$ and $n-m$ external legs with momenta $\vec k_{m+1}, \ldots, \vec k_n$, in the limit $q/k_i \to 0$. Expanding the exponential at a given order in $P_0(p)$ yields the number of soft lines exchanged. The integral in $\rmd^3 x$ ensures that  the sum of the internal momenta is $\vec q$.

Equation \eqref{soft_lines_final} can be easily generalized to consider the case where more than two sums of momenta become small, i.e.~when soft internal lines are exchanged between more than two hard-modes diagrams. The conclusion is always the same: soft internal lines do not contribute to equal time correlators at order $\propto k^2 \int \rmd p P_0(p) $. Again, this statement is very general irrespectively of the assumption about the short scales.

As a concrete example, let us consider the case $m=2, n=4$, i.e.~a 4-point function in the collapsed limit $|\vec k_1 + \vec k_2 |\ll k_1, k_2$, and the exchange of a single soft line. In this case, expanding the exponential at first order in $P_0(p)$, the above equation yields
\be
\label{t_wsl}
\begin{split}
\langle\delta_{\vec k_1}(\eta_1)& \delta_{\vec k_2}(\eta_2) \delta_{\vec k_{3}}(\eta_3) \delta_{\vec k_4}(\eta_4) \rangle_c' \\ &
\approx - \, \langle\delta(\vec k_1,\eta_1) \delta(\vec k_2,\eta_2) \rangle ' \langle \delta(\vec k_{3},\eta_{3}) \cdots \delta(\vec k_4,\eta_4)\rangle'\\
&\times   \int^\Lambda \rmd ^3  p \big( D(\eta_1) - D(\eta_2) \big) \frac{\vec k_1 \cdot \vec{p}}{p^2} \, \big(  D(\eta_3)  - D(\eta_4) \big) \frac{\vec k_3 \cdot \vec{p}}{p^2} \,  P_0(p) \delta_D(\vec p - \vec k_1 - \vec k_2)\,,
\end{split}
\ee
where we have considered only the connected diagram and, for simplicity, we are neglecting soft loops attached to each lines.
To compare with perturbation theory, we need to compute the tree-level exchange diagram. The contribution from taking $\vec k_1$ and $\vec k_3$ at second order yields 
\be
T_{2121} \approx - 4 D(\eta_1) D(\eta_3) P_0(|\vec k_1+\vec k_2|) 
\frac{\vec k_1 \cdot (\vec k_1 + \vec k_2 )}{  2 |\vec k_1 + \vec k_2|^2} \frac{\vec k_3 \cdot (\vec k_1 + \vec k_2 )}{ 2  |\vec k_1 + \vec k_2|^2} 
\langle \delta_{\vec k_1}(\eta_1) \delta_{\vec k_2}(\eta_2)\rangle'  \langle \delta_{\vec k_3}(\eta_3) \delta_{\vec k_4}(\eta_4)\rangle'   \; ,
\ee
and summing up the other permutations lead to
\be
\begin{split}
\langle\delta_{\vec k_1}(\eta_1)& \delta_{\vec k_2}(\eta_2) \delta_{\vec k_3}(\eta_3) \delta_{\vec k_4}(\eta_4)\rangle_c' \approx -  \big(D(\eta_1)-D(\eta_2)\big) \big( D(\eta_3) - D(\eta_4) \big) P_0(|\vec k_1+\vec k_2|) \\ & \times
\frac{\vec k_1 \cdot (\vec k_1 + \vec k_2 )}{  |\vec k_1 + \vec k_2|^2} \frac{\vec k_3 \cdot (\vec k_1 + \vec k_2 )}{ |\vec k_1 + \vec k_2|^2} 
\langle \delta_{\vec k_1}(\eta_1) \delta_{\vec k_2}(\eta_2)\rangle'  \langle \delta_{\vec k_3}(\eta_3) \delta_{\vec k_4}(\eta_4)\rangle'   \;,
\end{split}
\ee
which confirms eq.~\eqref{t_wsl}. 
One can easily extend this check to the case of several soft-lines.

\def\dg{\delta^{(g)}}

\section{\label{sec:redshift}Going to redshift space}

The derivation of the consistency relations has been done in real space, but the galaxy distribution will of course be observed in redshift space. It is thus natural to ask if it is possible to write relations directly in terms of redshift space correlation function. Before doing so, let us stress that it will be difficult---if not impossible---to measure  consistency relations at different times. To see the effect of the long mode, one would like to measure at quite different redshifts the short-scale correlation function at a spatial distance which is much smaller than Hubble. This is of course impossible since we can only observe objects on our past lightcone. This implies that, although one can check the consistency relations at different times in simulations, for real data we will have to stick to correlation functions at the same time. Given that the consistency relations vanish at equal time, their main phenomenological interest will be to look for their possible violations, which would indicate that one of the assumptions does not hold. This would represent a detection of either multi-field inflation or violation of the equivalence principle (or both!)

The mapping between real space $\vec x$ and redshift space $\vec s$ in the plane-parallel approximation is given by
\be
\vec s = \vec x + \frac{v_z}{\HH} \hat z\;, \label{rs-rs}
\ee
where $\hat z$ is the direction of the line of sight, $v_z \equiv \vec v \cdot \hat z$, and $\vec v$ is the peculiar velocity. Also the relation between $z$ and $\eta$ receives corrections due to peculiar velocities. These corrections are small for sufficiently distant objects for which $v \ll H x$. Notice that we do not assume that the peculiar velocity is a function of the position $\vec x$ since this holds only in the single-stream approximation, which breaks down for virialized objects on small scales \cite{Seljak:2011tx,Vlah:2012ni}. 

The derivation of the consistency relations follows closely what we did in real space, once we observe that also in redshift space the long mode induces a (time-dependent) translation. Indeed we have
\begin{align}
\vec x &\to \vec x  +  D \, \vec \nabla \Phi_{0,L}\;, \label{traslation} \\
\vec v & \to \vec  v+ f \HH D \, \vec  \nabla \Phi_{0,L} \label{vL} \;,
\end{align}
where $D(\eta)$ is the growth factor, $f(\eta) \equiv d \ln D/d \ln a$ is the growth rate and $\vec \nabla \Phi_{0,L}$ a homogenous gradient of the initial gravitational potential $\Phi_{0,L}$, related to $\delta_0$ defined in eq.~\eqref{deltax2delta} by $\nabla^2 \Phi_{0,L} = \delta_{0,L}$. 
This corresponds to a redshift space translation 
\begin{align}
\vec{\tilde s} &= \vec s + \delta \vec s\;, \\
\delta \vec s & \equiv D \, ( \vec \nabla \Phi_{0,L} + f  \nabla_z \Phi_{0,L} \hat z) \label{deltas}\;,
\end{align}
where we have applied to eq.~\eqref{rs-rs}  a spatial translation of the real-space coordinates and a  shift of the peculiar velocity along the line of sight, respectively eqs.~\eqref{traslation} and \eqref{vL}.
As in real space, we can thus conclude that a redshift-space correlation function in the presence of a long mode $\Phi_L$ is the same as the correlation function in the absence of the long mode but in {\em translated} redshift-space coordinates:\be
\begin{split}
\langle \dgs(\vec s_1,\eta_1) \cdots \dgs(\vec s_n,\eta_n) | {\Phi_L}\rangle &\approx \langle \dgs(\vec{\tilde  s}_1,\eta_1) \cdots \dgs(\vec{\tilde  s}_n,\eta_n) \rangle\; \\
& = \sum_a \delta \vec s_a  \langle \dgs(\vec s_1,\eta_1) \cdots \vec \nabla_a \dgs(\vec s_a,\eta_a) \cdots \dgs(\vec s_n,\eta_n) \rangle \;, \label{expansion}
\end{split}
\ee
where $\delta \vec s_a  \equiv D_a \, ( \vec \nabla_a \Phi_{0,L} + f_a  \nabla_{a,z} \Phi_{0,L} \hat z) $. To show this notice that the density in redshift space can be written in terms of the real-space distribution function \cite{Seljak:2011tx,Vlah:2012ni}
\be
\rho_s(\vec s) = m a^{-3} \int \rmd^3 p \; f \left(\vec s - \frac{v_z}{\cal H} \hat z, \vec p\right) \;,
\ee
where $m$ is the mass of the particles and $\vec p$ is the physical momentum. The statistical properties of $\rho_s(\vec s)$ in the presence of the long mode are inherited by its expression in real space
\be
\rho_s(\vec s)_{\Phi_L} =  \frac{m}{a^3} \int \rmd^3 p \; f \left(\vec s - \frac{v_z}{\cal H} \hat z +\delta\vec x, \vec p +a m \delta \vec v\right) = \frac{m}{a^3} \int \rmd^3 p' \; f \left(\vec s - \frac{v_z- \delta v_z}{\cal H} \hat z +\delta\vec x, \vec {p'}\right) = \rho_s(\vec s + \delta \vec s),
\ee 
where $\delta x$ and $\delta \vec v$ are given by eqs.~\eqref{traslation} and \eqref{vL}.

Again this statement can be directly applied to the galaxy distribution and it thus includes the bias with respect to the dark matter distribution.
Notice that in the plane-parallel approximation redshift space is still translationally invariant (although it is not rotationally invariant, since the line-of-sight is a preferred direction): correlation function only depends on the distance between points. This implies that the consistency relations will be zero when the short modes are taken at equal time, since the common translation does not change distances.

In the Fourier space conjugate to redshift space, eq.~\eqref{expansion} becomes
\be
\langle  {\Phi_{0}}({\vec q}) \dgs_{\vec k_1}(\eta_1) \cdots \dgs_{\vec k_n}(\eta_n) \rangle \approx P_\Phi (q) \sum_a D(\eta_a) \big[  \vec q \cdot \vec k_a + f(\eta_a) q_z \, k_{a,z} \big]  
\langle  \dgs_{\vec k_1}(\eta_1) \cdots \dgs_{\vec k_n}(\eta_n) \rangle \;.
\ee
By using for the long mode the linear relation between the density contrast in redshift space $\delta$ and the gravitational potential $\Phi$, i.e.
\be
\dgs({\vec q},\eta) = - (b_1+ f \mu_{\vec q}^2) D(\eta) q^2 \Phi_0 ({\vec q}) \;,
\ee
where $b_1$ is a linear bias parameter between galaxies and dark matter and $\mu_{\vec k} \equiv \vec k \cdot \hat z/k$, the consistency relation above becomes
\be
\label{CR_rs}
\begin{split}
\langle  \dgs_{\vec q}(\eta) \dgs_{\vec k_1}(\eta_1) \cdots \dgs_{\vec k_n}(\eta_n) \rangle \approx &  - \frac{P_{g,s}(q,\eta)}{b_1+ f \mu_{\vec q}^2} \sum_a \frac{D(\eta_a)}{D(\eta)} \frac{k_a}{q} \big[  \hat q \cdot \hat k_a + f(\eta_a) \mu_{\vec q} \, \mu_{\vec k_a} \big] \\ &\times \langle  \dgs_{\vec k_1}(\eta_1) \cdots \dgs_{\vec k_n}(\eta_n) \rangle \;.
\end{split}
\ee

We can check that this relation holds in perturbative calculation of redshift space distortions. The redshift space bispectrum reads \cite{Bernardeau:2001qr}
\be
\begin{split}
\langle  \dgs_{\vec q}(\eta) \dgs_{\vec k_1}(\eta_1) \dgs_{\vec k_2}(\eta_2) \rangle' & = \\ 2 Z_2(-\vec q, -\vec k_2;\eta_1) Z_1(\vec q;\eta) Z_1(\vec k_2;\eta_2)   &\langle  \delta({\vec q}, \eta) \delta({-\vec q},\eta_1) \rangle' \langle \delta({\vec k_1},\eta_1) \delta({\vec k_2}, \eta_2) \rangle'  \text{ + cyclic}\;,
\end{split}
\ee
where
\be
\begin{split}
Z_1(\vec k;\eta) & \equiv (b_1 + f \mu_{\vec k}^2) \;, \\
Z_2(\vec k_a,\vec k_b;\eta) & \equiv b_1 F_2(\vec k_a,\vec k_b) +f \mu_{\vec k}^2 G_2(\vec k_a,\vec k_b) + \frac{f \mu_{\vec k} k}{2} \left[\frac{\mu_{\vec k_a}}{k_a}(b_1+f\mu_{\vec k_b}^2)+ \frac{\mu_{\vec k_b}}{k_b}(b_1+f\mu_{\vec k_a}^2)\right] +\frac{b_2}{2} \;.
\end{split}
\ee
Here $b_1$ and $b_2$ are the linear and non-linear bias parameters and $F_2$ and $G_2$ are the standard second-order perturbation kernels for density and velocity respectively \cite{Bernardeau:2001qr}.
In the limit $q \to 0$ we have
\be
2 Z_2 (-\vec q, -\vec k_2;\eta_1) \approx (b_1 + f_1 \mu_{\vec k_2}^2) \frac{\vec q \cdot \vec k_2}{q^2} + (b_1 + f_1 \mu_{\vec k_2}^2) f_1 \frac{k_2}{q} \mu_{\vec q} \mu_{\vec k_2} \;.
\ee
This gives
\be
\begin{split}
\langle  \dgs_{\vec q}(\eta) & \dgs_{\vec k_1}(\eta_1) \dgs_{\vec k_2}(\eta_2) \rangle'  \\ & \approx \frac{P_{g,s}(q,\eta)}{b_1+ f \mu_{\vec q}^2} \frac{D(\eta_1)}{D(\eta)}  (b_1 + f_1 \mu_{\vec k_2}^2) \left( \frac{\vec q \cdot \vec k_2}{q^2} +  f_1 \frac{k_2}{q} \mu_{\vec q} \mu_{\vec k_2} \right)   Z_1(\vec k_2;\eta_2)  \langle  \delta({\vec k_1},\eta_1) \delta({\vec k_2}, \eta_2) \rangle' \\ & \approx -\frac{P_{g,s}(q,\eta)}{b_1+ f \mu_{\vec q}^2} \frac{D(\eta_1)}{D(\eta)}   \frac{k_1}{q} (\hat q \cdot \hat k_1 +  f_1  \mu_{\vec q} \mu_{\vec k_1})   \langle  \delta({\vec k_1},\eta_1) \delta({\vec k_2}, \eta_2) \rangle'  \text{  + $(1 \leftrightarrow 2)$}\;.
\end{split}
\ee
The consistency relation is satisfied.

As in real space, it is possible to derive a resummed version of eq.~\eqref{CR_rs}. The translation in redshift space introduces a factor
\be
\exp \Big[ i  \sum_a \vec k_a \cdot \delta \vec s( \vec y,\eta_a) \Big] =\exp \Big[  \int^\Lambda\frac{\rmd^3p}{(2\pi)^3} \sum_a D(\eta_a)\left( \vec p \cdot \vec k_a + f(\eta_a) p_z \, k_{a,z}\right)  \,e^{i  \vec p \cdot \vec{y}}\Phi_{0}(\vec p) \Big]\;
\ee
in the correlation functions. It is then straightforward to show that, as in Sec.~\ref{sec:resum}, the consistency relation in redshift space eq.~\eqref{CR_rs} remains the same even when the effect of all soft modes is resummed.
%\be
%\begin{split}
%\langle  \dgs_{\vec q}(\eta) &\dgs_{\vec k_1}(\eta_1) \cdots \dgs_{\vec k_n}(\eta_n) \rangle \approx  - \frac{P_{g,s}(q,\eta)}{b_1+ f \mu_{\vec q}^2} \sum_a \frac{D(\eta_a)}{D(\eta)} \frac{k_a}{q} \big[  \hat q \cdot \hat k_a + f(\eta_a) \mu_{\vec q} \, \mu_{\vec k_a} \big]  \\
%&  \times \exp \bigg[ {- \frac12 \int \frac{\rmd^3 p}{(2\pi)^3}  \bigg( \sum_a D(\eta_a) \frac{\vec k_a \cdot \vec{p}+ f(\eta_a) p_z \, k_{a,z}} {p^2} \, \bigg)^2 P_0(p)} \bigg] 
%\langle  \dgs_{\vec k_1}(\eta_1) \cdots \dgs_{\vec k_n}(\eta_n) \rangle\; . \label{CRRedshift}
%\end{split}
%\ee
Moreover, using the same procedures developed in the previous section, one can easily extend the consistency relations with multiple soft legs, softs loops and soft internal lines to redshift space.

\section{Conclusions}
In this paper we showed that one can have a complete control of soft modes at any order in $\frac{k}{q} \cdot \delta_q$. The known cancellation of these effects for equal time correlators \cite{Jain:1995kx,Scoccimarro:1995if,Bernardeau:2011vy,Bernardeau:2012aq,Blas:2013bpa,Carrasco:2013sva} is now on more general grounds:  it is physically a consequence of the equivalence principle and the lack of statistical correlation between long and short modes, which holds in single-field inflation. Therefore this cancellation is very robust and holds beyond the single-stream approximation, and including the effects of baryons on short scales. These regimes are beyond the usual arguments based on perturbation theory.  Moreover, we now know exactly what is the effect of soft modes on correlators at different times.
To make contact with observations one has to understand if the consistency relations can be written directly in redshift space. We showed that this is the case, without adding any assumption about the short modes: for example one does not need to assume the single-stream approximation, which breaks down on short scales.

Besides the theoretical interest of these results, the main conclusion for observations is that a detection in the squeezed limit of a $1/q$ behaviour at equal time would be a robust detection of either multi-field inflation or a violation of the equivalence principle. The next step is to evaluate how constraining measurements will be for explicit models that do not respect equivalence principle, taking into account that in the data one is obviously limited in the hierarchy between $k$ and $q$. We will come back to this in a future publication \cite{Creminelli:2013nua}.

\subsection*{Acknowledgements}
While finishing this paper reference \cite{Peloso:2013spa} appeared. There is no disagreement with our results: in particular, we both agree that a violation of the EP implies a breaking of the consistency relations in the form of eq.~\eqref{CR}. We thank M.~Peloso and M.~Pietroni for discussions. We acknowledge related work by A.~Kehagias, J.~Nore\~na, H.~Perrier and A.~Riotto: where comparison is possible, the results agree.
It is a pleasure to thank V.~Desjacques, R.~Scoccimarro and M.~Zaldarriaga for useful discussions, and the anonymous referee for useful comments. JG and FV acknowledge partial support by the ANR {\it Chaire d'excellence} CMBsecond ANR-09-CEXC-004-01.

\end{document}